\documentclass[%
aip,
amsmath,amssymb,
reprint,%
]{revtex4-1}

\usepackage{graphicx}               
\usepackage{dcolumn}                
\usepackage{bm}                     

\usepackage[utf8]{inputenc}
\usepackage[T1]{fontenc}
\usepackage{amsmath}
\usepackage{amssymb}
\usepackage{braket}
\usepackage{longtable}
\usepackage{siunitx}
\usepackage{etoolbox}
\usepackage{gensymb}
\usepackage{bbold}

\usepackage{natbib}

\usepackage{enumitem}

\makeatletter
\def\@email#1#2{%
 \endgroup
 \patchcmd{\titleblock@produce}
  {\frontmatter@RRAPformat}
  {\frontmatter@RRAPformat{\produce@RRAP{*#1\href{mailto:#2}{#2}}}\frontmatter@RRAPformat}
  {}{}
}%
\makeatother

\DeclareSIUnit[number-unit-product = {\,}]\cal{cal}
\DeclareSIUnit[number-unit-product = {\,}]\mol{mol}

\usepackage{xcolor}
\usepackage[normalem]{ulem}

\newcommand{\trial}{\mathrm{T}}
\newcommand{\init}{\mathrm{I}}
\newcommand{\Heff}{\mathcal{H}}

\begin{document}

\preprint{AIP/123-QED}

\title[]{Towards Large-Scale AFQMC Calculations: Large Time Step Auxiliary-Field Quantum Monte Carlo}

\author{Zoran Sukurma}
\affiliation{University of Vienna, Faculty of Physics and Center for Computational Materials Science, Kolingasse 14-16, A-1090 Vienna, Austria}
\affiliation{University of Vienna, Faculty of Physics \& Vienna Doctoral School in Physics,  Boltzmanngasse 5, A-1090 Vienna, Austria}
\author{Martin Schlipf}
\affiliation{VASP Software GmbH, Sensengasse 8, 1090 Vienna, Austria}
\author{Moritz Humer}
\affiliation{University of Vienna, Faculty of Physics and Center for Computational Materials Science, Kolingasse 14-16, A-1090 Vienna, Austria}
\affiliation{University of Vienna, Faculty of Physics \& Vienna Doctoral School in Physics,  Boltzmanngasse 5, A-1090 Vienna, Austria}
\author{Amir Taheridehkordi}
\affiliation{University of Vienna, Faculty of Physics and Center for Computational Materials Science, Kolingasse 14-16, A-1090 Vienna, Austria}
\author{Georg Kresse}
\affiliation{University of Vienna, Faculty of Physics and Center for Computational Materials Science, Kolingasse 14-16, A-1090 Vienna, Austria}
\affiliation{VASP Software GmbH, Sensengasse 8, 1090 Vienna, Austria}

\date{\today}

\begin{abstract}
We report modifications of the ph-AFQMC algorithm that allow the use of large time steps and reliable time step extrapolation.
Our modified algorithm eliminates size-consistency errors present in the standard algorithm when large time steps are employed.
We investigate various methods to approximate the exponential of the one-body operator within the AFQMC framework, distinctly demonstrating the superiority of Krylov methods over the conventional Taylor expansion.
We assess various propagators within AFQMC and demonstrate that the Split-2 propagator is the optimal method, exhibiting the smallest time-step errors.
For the HEAT set molecules, the time-step extrapolated energies deviate on average by only \qty{0.19}{\kilo \cal / \mol} from the accurate small time-step energies.
For small water clusters, we obtain accurate complete basis-set binding energies using time-step extrapolation with a mean absolute error of \qty{0.07}{\kilo \cal / \mol} compared to CCSD(T).
Using large time-step ph-AFQMC for the N$_2$ dimer, we show that accurate bond lengths can be obtained while reducing CPU time by an order of magnitude. 
\end{abstract}

\maketitle

\section{Introduction}
\label{sec:intro}
The accurate description of electron-electron correlation in large systems remains a fundamental challenge in solid-state physics and quantum chemistry.
When beyond DFT accuracy is needed, one commonly resorts to coupled-cluster singles, doubles, and perturbative triples (CCSD(T))\cite{Cizek1966,CizekPaldus1971} or fixed-node diffusion Monte Carlo (FN-DMC).\cite{Anderson1976DMC,Ceperley1977DMC} 
Over the past two decades, several accurate correlation-consistent methods have emerged as promising tools for achieving nearly chemical accuracy:
density-matrix renormalization group,\cite{White1999DMRG,Schollwock2005DMRG,Chan2011DMRGRev}
full configuration-interaction quantum Monte Carlo,\cite{Booth2009FCIQMC,Cleland2010IFCIQMC,Booth2013FCIQM,cleland2012taming,Ghanem2020ASFCIQMC} 
heat-bath configuration interaction,\cite{PetruzieloSPMC2012,Umrigar2016HeatBath,HolmesHCIFock2016,Sharma2017HeatBath}
adaptive sampling configuration interaction,\cite{TubmanASCI2016,TubmanASCIED2020,LevineASCICASSCF2020}
iterative CI,\cite{LiuiCI2016,ZhangiCI2020}
and many-body expansion of full configuration interaction (MBE-FCI). \cite{EriksenMBEFCI2017,EriksenMBEFCIWeak2018,EriksenMBEFCIStrong2019,EriksenGenMBEFCI2019}
Extensive tests on small systems show that these methods yield nearly exact FCI solutions.
As a result, they serve as valuable benchmark tools or high-level solvers e.g. in embedding schemes.
However, their application is currently limited to modest system sizes due to high computational costs and other inherent limitations.
Besides the mentioned methods, phaseless auxiliary-field quantum Monte Carlo (ph-AFQMC)\cite{Zhang2003Phaseless} is one of the most promising alternatives to both CCSD(T) and FN-DMC because of its favorable scaling with system size and because it is readily applicable to molecular and extended systems.

\v{C}\'{\i}\v{z}ek and Paldus \cite{Cizek1966,CizekPaldus1971} introduced the
coupled-cluster theory 50 years ago, and since then, it has found extensive use in correlation-consistent calculations.
The \emph{gold standard} of quantum chemistry---coupled-cluster singles, doubles, and perturbative triples (CCSD(T))\cite{RaghavachariCCSDT1989,Helgaker2000QC,Bartlett2007CC}--- has been carefully benchmarked on a wide range of small molecules, achieving nearly chemical accuracy. \cite{Bartlett2007CC,Feller2001CCSDT}
The Gr\"uneis and Chan groups have successfully applied CCSD(T) to solids routinely treating a few hundred electrons.\cite{GruberCC2018,Gruneis2017perspective,Gruneis2019CCSDT,IrmlerFocalPoint2021,McClain_ccsdt_solids_2017}
However, CCSD(T) is often not accurate enough, especially in the presence of strong static correlation and spin contamination. \cite{Bartlett2007CC}
For instance, the total energy of the CN radical in a double-zeta basis set converges to within \qty{0.3}{\kilo \cal / \mol} only at the CCSDTQP level of theory.\cite{Bomble2005Heat}
More importantly, the CCSD(T) theory already exhibits adverse computational scaling ($\mathcal{O}(N^7)$) and higher coupled-cluster expansions, such as CCSDTQP, are accessible only for the smallest systems (20 electrons in 100 orbitals).
Similar to stochastic and selected versions of CI, analogous coupled-cluster variants have been proposed.
These aim to alleviate the computational scaling of higher coupled-cluster expansions and enhance accuracy beyond CCSD(T).\cite{ThomCC2010,FilipMRCC2019,FilipUCC2020,ShenCCPQ2012,Deustua2017,Chakraborty2022}
More extensive benchmark data for these methods is necessary to judge the performance and accuracy relative to CCSD(T).

Fixed-node diffusion Monte Carlo\cite{Anderson1976DMC,Ceperley1977DMC,Foulkes2001QMC} (FN-DMC) is a commonly used method for studying solids and molecules containing up to 1000 electrons.
It uses a random walk in real space to sample the Hamiltonian expectation values and scales as $\mathcal{O}(N^3)-\mathcal{O}(N^4)$ with system size. 
DMC requires accurate trial wavefunctions to guide the walker density to the important positions in real space.
Typically, Jastrow factors\cite{JastrowOrig1955,JastrowSchmidt1990,JastrowDrummond2004} on top of the Slater determinant impose cusp conditions.
While Jastrow factors yield faster convergence to the complete basis set (CBS) limit, FN-DMC becomes comparable to other quantum chemistry methods only in the CBS limit.
Several groups showed that the mean absolute error of the FN-DMC with a single Slater-Jastrow wavefunction is approximately \qty{3}{\kilo \cal / \mol}.\cite{GrossmanBench2002,Nemec2010DMC,PetruzieloDMC2012}
Petruzielo \emph{et al.} \cite{PetruzieloDMC2012} demonstrated that a variational optimization of orbitals within the Slater-Jastrow wavefunction reduces the error to \qty{2.1}{\kilo \cal / \mol}.
Furthermore, adding multiple Slater determinants from small complete active spaces reduces the mean absolute error to \qty{1.2}{\kilo \cal / \mol} for the G2 set.
To mitigate the fermionic sign problem, DMC relies on the fixed-node approximation.
This constrains the nodal structure of the exact many-body ground-state wavefunction to match the nodes of the trial wavefunction.
However, the trial wavefunction is the primary source of errors in DMC.\cite{Anderson1976DMC,LuechowFN2001}
More accurate trial wavefunctions such as backflow wavefunctions,\cite{BackflowLopezRios2006} geminal wavefunctions,\cite{GeminalCasula2003} pfaffians,\cite{PfaffianBajdich2006} or recently introduced deep learning models,\cite{Pfau2020Ferminet,Scherbela2022DeepErwin}
provide a route for further improvements of the method.

In this work, we will focus on the phaseless auxiliary-field quantum Monte Carlo (ph-AFQMC).
Although originally developed as the path integral MC algorithm, \cite{BlankenbeclerAFQMC1981,ScalapinoAFQMC1981,Sugiyama1986AFQMC,Sorella1989AFQMC} it experienced a renaissance with the introduction of open-ended random walks and the phaseless approximation.\cite{Zhang2003Phaseless}
Since then, numerous applications for molecules \cite{Al-Saidi2006Gaussian,PurwantoSpinCont2008, Purwanto2009Excited,Purwanto2105ChromiumDimer,Al-SaidiDEl2006,Al-SaidiH2007,Al-Saidi2007BondBreaking,Motta2017Hchain,Borda2019NONSD,sukurmaHEAT2023} and solids \cite{Suewattana2007Solids,PurwantoSiTin2009,MaAFQMCExc2013,LeeHEG2019,Amir2023} have confirmed that ph-AFQMC is a promising alternative to CCSD(T) and FN-DMC.
It has also been successfully applied to excited states \cite{Purwanto2009Excited,MaAFQMCExc2013} and at finite temperature.\cite{ZhangFTAFQMC1999,ZhangFTAFQMC2000,Liu2018AbInitio,GilberthFastFTAFQMC2021,LeeHEGFT2021}
ph-AFQMC exhibits low polynomial scaling $(\mathcal{O}(N^3)-\mathcal{O}(N^4))$ with system size that can even be reduced to cubic scaling by utilizing hypertensor contractions\cite{Motta2109-AFQMC-ED,Malone2019AFQMC-DF}, resolution of identity,\cite{Lee2020AFQMCreduced} or employing a plane wave basis.\cite{Suewattana2007Solids}
The \emph{phaseless} approximation addresses the fermionic sign problem in ph-AFQMC, however, it also introduces systematic errors that are challenging to control.\cite{Zhang2003Phaseless,Motta2018AFQMCREVIEW}
Mahajan \emph{et al.} developed a fast method for the evaluation of the local energies for multi-determinantal trial wavefunctions and demonstrated that the phaseless approximation can be systematically improved using better trial wavefunctions.\cite{Mahajan2021TamingSignProblem,Mahajan2022AFQMCsCI,Maloneipie2023}
Recently, there have been attempts to directly enhance the phaseless approximation.\cite{sukurmaHEAT2023,XiaoAFQMCconstr2023,WeberAFQMCconstr2023}
 
In this work, we focus on reducing the computational prefactor of ph-AFQMC by utilizing large time steps.
Larger time steps facilitate faster equilibration, diminish the autocorrelation between subsequent local energy evaluations, and thereby significantly reduce the computational cost to achieve fixed statistical errors (see Sec.~\ref{sec:timestep}).
Therefore, every projector QMC method employs the largest time step not compromising the accuracy.
For typical systems, ph-AFQMC utilizes time steps around \qty{e-3}{\hartree^{-1}} without significant time-step errors. \cite{Motta2018AFQMCREVIEW,Shi2021RecentDevAFQMC,sukurmaHEAT2023}
In contrast, FN-DMC permits much larger time steps up to \qty{e-1}{\hartree^{-1}}.\cite{Umrigar1993POP,Zen2016Boosting,umrigar2023timestep}
In FCIQMC, the time step is limited by the maximal excitation energy.\cite{Booth2009FCIQMC}

We modify the AFQMC algorithm to enable size-consistent simulations with large time steps.
To eliminate the remaining time-step errors, we investigate various approaches to compute the action of the one-body propagator on a Slater determinant.
The optimal method minimizes the required number of $\hat H \ket{\Psi}$ operations for a set of representative systems.
Further, we adapt various propagators within the AFQMC framework and analyze their corresponding leading error terms. 
We show that the time-step errors agree with the theoretical predictions and that a reliable and robust time-step extrapolation to a zero time-step limit is possible.
Finally, we show the accuracy of time-step extrapolated AFQMC energies for a variety of molecular systems and Gaussian basis sets.

The remainder of the paper is organized as follows:
in Sec.~\ref{sec:afqmc_rev} we briefly summarize the ph-AFQMC method, and in Sec.~\ref{sec:timestep} we explore the advantages associated with the use of large time steps.
We investigate different propagators within the ph-AFQMC framework in Sec.~\ref{sec:integrators}, followed by a comparison of methods to compute the exponential of the one-body operator in Sec.~\ref{sec:expm}).
Then, practical recipes for ph-AFQMC simulations with large time steps are presented in Sec.~\ref{sec:modalgo}.
We validate our method through various applications in Sec.~\ref{sec:results}.
Finally, Sec.~\ref{sec:conclusion} summarizes our findings and outlines the perspectives of AFQMC in the simulation of real materials.

\section{Theoretical Background}

\subsection{AFQMC Review} 
\label{sec:afqmc_rev}
In this section, we will briefly introduce the ph-AFQMC formalism.
For a more detailed overview of the theory, we recommend the reviews in Refs.~\onlinecite{Zhang2003Phaseless,Zhang2013Juelich,Motta2018AFQMCREVIEW}.

Auxiliary-field quantum Monte Carlo projects an initial state $\ket{\Psi_{\init}}$ onto the the exact ground-state wavefunction $\ket{\Phi}$
using the imaginary time Schr\"odinger equation
\begin{equation}
    \label{eq:imag_prop}
    \ket{\Phi} = \lim_{n \to \infty} \left[ e^{-\tau \left( \hat H - E_0 \right) }  \right]^{n} \ket{\Psi_{\init}},
\end{equation}
where $\tau$ is the imaginary time step, $\hat H$ is the many-body Hamiltonian, and $E_0$ is the best estimate of the ground-state energy.
For practical reasons, the initial state $\ket{\Psi_{\init}}$ is usually a single Slater determinant formed from Hartree-Fock or Kohn-Sham orbitals.

AFQMC is suitable to treat Hamiltonians of the form
\begin{equation}
\label{eq:mbham}
    \hat H   = \sum_{pq} h_{pq} \; \hat a_{p}^{\vphantom{\dagger}\dagger} \hat a_{q} \ + \ \frac{1}{2} \sum_{pqrs} V_{pqrs} \; \hat a_{p}^{\vphantom{\dagger}\dagger} \hat a_{q}^{\vphantom{\dagger}\dagger} \hat a_{s}  \hat a_{r},
\end{equation}
where $\hat a_{p}^{\dagger}$ and $\hat a_{q}$ are fermionic creation and annihilation operators, respectively and the $h_{pq}$ matrix elements represent the one-body Hamiltonian $\hat H_1$.
The two-body terms $V_{pqrs}$ are usually the electron repulsion integrals $\braket{pq|rs}$ and must be positive definite to decompose them to
\begin{equation}
    \braket{pq|rs} = \sum_g L_{g,pr} L_{g,qs}
\end{equation}
using a spectral decomposition,\cite{Zhang2003Phaseless} an iterative Cholesky decomposition,\cite{Beebe1977CholDec,Koch2003CholDec,Motta2018AFQMCREVIEW} density fitting techniques,\cite{RendellDFCCSDT1994,DunlapDF2000,WernerDFMP22003} or plane waves.\cite{Zhang2005Solids,Suewattana2007Solids,Amir2023}
The indices $p$, $q$, $r$, and $s$ go over $N$ basis functions, while the index $g$ goes over the size of the decomposition $N_g$.
In the Gaussian basis, we usually have $N_g \approx 10N$.
In this work, we use a Cholesky decomposition and refer to the $L_{g,pq}$ tensors as Cholesky vectors.
Specifically, we opt for the iterative Cholesky procedure because a single threshold parameter controls the accuracy of the decomposition and the algorithm stops when reaching the desired accuracy, i.e., does not compute superfluous $L_{g,pq}$ tensors.

To treat the two-body Hamiltonian in Eq.~\eqref{eq:imag_prop}, AFQMC employes the Hubbard-Stratonovich transformation\cite{Hubbard1959,Stratonovich1957}
\begin{equation}
\label{eq:hstrafo}
    e^{-\frac{\tau}{2} \sum_g \hat L_{g}^{2}} = \int \text{d}x^{N_g} \; p(\textbf{x}) \; e^{i \sqrt{\tau} \sum_g x_g \hat L_g} + \mathcal{O}(\tau^2), 
\end{equation}
to map the interacting system onto a non-interacting one coupled to a fluctuating random field.
$p(\textbf{x})$ denotes an $N_g$-dimensional standard Gaussian probability density with the random numbers $x_g$, while $\hat L_g$ is the one-particle operator formed from Cholesky vectors
\begin{equation}
    \hat L_g = \sum_{pq} L_{g,pq} \; \hat a_{p}^{\vphantom{\dagger}\dagger} \hat a_{q}~.
\end{equation}
Equation \eqref{eq:hstrafo} is the core of the method and the integration is performed using Monte Carlo sampling.
To realize an efficient Monte Carlo sampling, we approximate the exact many-body ground state wavefunction at the time step $k$ as an ensemble of $N_w$ walkers
\begin{equation}
\label{eq:wavefun_rep}
    \ket{\Phi_k} = \Bigl(\sum_{w} W_{k}^{w} e^{i \theta_{k}^{w}}\Bigr)^{-1} \sum_{w} W_{k}^{w} e^{i \theta_{k}^{w}} \frac{\ket{\Psi_{k}^{w}}}{\braket{\Psi_{\trial} | \Psi_{k}^{w}}}.
\end{equation}
We represent each walker by a single Slater determinant $\ket{\Psi_k^w}$, a real-valued weight $W_k^w$, and the phase $\theta_k^w$.
The walkers are initialized as
\begin{align}
    \ket{\Psi_{0}^w} & = \ket{\Psi_{\init}};
    &
    W_{0}^w e^{i\theta_0^w} & = \braket{\Psi_{\trial}|\Psi_{\init}},
\end{align}
where $\ket{\Psi_{\trial}}$ is the trial wavefunction.
In time step $k$, we approximate the integral in Eq.~\eqref{eq:hstrafo} by $N_w$ sets of random fields $x_g^w$ that form the effective interaction $\hat{\Heff}_{\mathrm{int}}^w$ with the matrix representation
\begin{equation}
\label{eq:heffTOT}
    \Heff_{\mathrm{int},pq}^{w} = - \frac{i}{\sqrt{\tau}} \sum_{g} (x_{g}^{w} - f_{g}^{w})  (L_{g,pq} - \bar L_{g} \delta_{pq}),
\end{equation}
where the mean-field shift $\bar L_g$ and the force bias $f_g^w$ minimize the variance of the sampling
\begin{equation}
\label{eq:force_shift}
\begin{split}
    \bar L_g & = \braket{\hat L_g}_{\trial} = \frac{\braket{\Psi_{\trial}|\hat L_g | \Psi_{\trial}}}{\braket{\Psi_{\trial}| \Psi_{\trial}}}  \\
    f_{g}^{w} & = - i \sqrt{\tau} \ \frac{\braket{\Psi_{\trial} | \hat L_g - \bar L_g | \Psi^{w} }}{\braket{\Psi_{\trial} | \Psi^{w}}}.
\end{split}
\end{equation}
The reviews in Refs.~\onlinecite{Rom1997Shifted,Zhang2013Juelich,Motta2018AFQMCREVIEW} provide more details about variance reduction techniques within the AFQMC formalism.
The equations of motion for the walkers are
\begin{gather}
    \ket{\Psi_{k+1}^w} = \hat U(\tau; \hat H_1, \hat \Heff_{\mathrm{int}}) \ket{\Psi_{k}^w},    \label{eq:psiupdate}    \\
    W_{k+1}^w e^{i \theta_{k+1}^w} = W_{k}^w e^{i \theta_{k}^w} \; e^{-\tau (E_{\textrm{H}}(\Psi_{k+1}^{w}) - E_0)},   \label{eq:wupdate}
\end{gather}
with the hybrid energy
\begin{equation}
    \label{eq:hyben}
    E_{\textrm{H}}(\Psi_{k+1}^{w}) = - \frac{1}{\tau} \log \Bigl( \frac{\braket{\Psi_{\trial} | \Psi_{k+1}^w}}{\braket{\Psi_{\trial} | \Psi_{k}^w}} I^{w} \Bigr),
\end{equation}
and the importance sampling factor
\begin{equation}
    I^{w} = \exp\Bigl[\sum_g x_g^w f_g^w - \frac{1}{2} f_g^w f_g^w\Bigr].
\end{equation}
The choice of the propagator $\hat U$ in Eq.~\eqref{eq:psiupdate} is thoroughly explored in Sec.~\ref{sec:integrators}.
Equations \eqref{eq:psiupdate} and \eqref{eq:wupdate} define the free-projection AFQMC (fp-AFQMC).
Although formally exact, fp-AFQMC suffers from the fermionic phase problem, where walkers can acquire any phase $e^{i\theta}$ with $\theta \in [0, 2\pi)$. 
It is equivalent to the sign problem in diffusion Monte Carlo except that for DMC walkers only $e^{i\theta} = \pm 1$ are possible.
As a consequence, the Monte Carlo estimator of the quantity $\braket{\Psi_{\trial}|\Phi}$ averages to zero. 
To circumvent this problem, Zhang \emph{et al.} \cite{Zhang2003Phaseless} proposed the \emph{phaseless approximation} (ph-AFQMC), i.e. a modification of the Eq.~\eqref{eq:wupdate} 
 \begin{gather}
    \begin{aligned}
    \label{eq:phaseless}
    W_{k+1}^w = W_{k}^w \: e^{-\tau (\mathrm{Re} E_{\textrm{H}} - E_0)} \ \text{max}\left( 0, \: \text{cos}(\Delta \theta) \right),
    \end{aligned}   \\
    \theta_{k}^w = 0,
\end{gather}   
where $\Delta \theta$ is the phase change in the overlap that one walker acquires after one time step
\begin{equation}
    \label{eq:theta}
    \Delta \theta = \mathrm{Im} \ \log \frac{\braket{\Psi_{\trial} | \Psi_{k+1}^w}}{\braket{\Psi_{\trial} | \Psi_{k}^w}}.
\end{equation}
The phaseless approximation keeps the algorithm stable, but it also introduces a systematic error known as the \emph{phaseless error}.
We investigated modifications to the phaseless approximation that significantly reduce the overcorrelation issues frequently encountered in ph-AFQMC, but they did not demonstrate systematic improvement compared to the standard phaseless approximation.\cite{sukurmaHEAT2023}
However, the phaseless errors can be systematically reduced by improving the trial wavefunctions\cite{Mahajan2021TamingSignProblem,Mahajan2022AFQMCsCI,Maloneipie2023} or by release constraint techniques.\cite{XiaoAFQMCconstr2023,WeberAFQMCconstr2023}

We perform $N$ Monte Carlo steps with the time step $\tau$ and $N$ energy measurements
\begin{align}
    W_k & = \sum_w W_k^w
    &
    E_k & =  \frac{\sum_w W_k^w E_{\mathrm{L}}(\Psi_k^w)}{W_k},
\end{align}
where $E_{\mathrm{L}}(\Psi_k^w)$ is the local energy of the walker $\Psi_k^w$
\begin{equation}
    \label{eq:eloc}
    E_{\mathrm{L}} = \frac{\braket{\Psi_{\trial} | \hat H | \Psi_k^w}}{\braket{\Psi_{\trial} | \Psi_k^w}}.
\end{equation}
The time average $\overline{E}$ and the variance $\sigma_E^2$ are defined as 
\begin{align}
    \overline{E} = & \frac{\sum_{k=N_{\mathrm{eq}}+1}^{N} W_k E_k}{\sum_{k=N_{\mathrm{eq}}+1}^{N} W_k},
    &
    \sigma_E^2 = & \overline{E^2} - \overline{E}^2,
\end{align}
where $\overline{E}$ is normally distributed around a mean $E$ with the variance
\begin{equation}
\label{eq:variance}
  \sigma_{\overline{E}}^2 = \frac{\sigma_{E}^2}{N} \kappa~.  
\end{equation}
The initial $N_{\mathrm{eq}}$ samples define the equilibration phase and are excluded from the average to avoid excited state mixing.
The variance in Eq.~\eqref{eq:variance} is scaled by a correlation length $\kappa$, representing the number of steps after which two samples $E_k$ and $E_{k+\kappa}$ are statistically independent.

\subsection{Computational Advantage of the Large Time-Step AFQMC}
\label{sec:timestep}
The first advantage of ph-AFQMC with large time steps is the reduction of the number of steps to equilibrate the systems.
We need to propagate each walker for $N_{\mathrm{eq}}$ steps so that the initial state $\ket{\Psi_\init}$ is projected to the ground state.
The number of required equilibration steps $N_\mathrm{eq}$ decreases like $1/\tau$ with the size of the time step.

The second advantage of large time-steps is the reduction of the autocorrelation between subsequent samples.
After taking a sample of the energy, one needs to propagate for $\kappa$ steps to avoid autocorrelation between the samples.
Ideally, we would increase the time step until $\kappa = 1$ so that we can sample the energy after every time step.

In practice, the benefits of large time steps are reduced by the fact that the energy does not have to be measured at each time step, if the samples are correlated.
However, other tasks that constitute the random walk (creation of the effective Hamiltonian, AFQMC propagation, and the force bias calculation) are still needed at each time step.
\cite{sukurmaHEAT2023}

\section{AFQMC Propagators}
\label{sec:integrators}
The exact many-body propagator is represented by
\begin{equation}
    \label{eq:exprop}
    \hat U(\tau) = e^{-\tau (\hat H_1 + \hat H_2)},
\end{equation}
where $\hat H_1$ and $\hat H_2$ are the one- and two-body terms defined in Eq.~\eqref{eq:mbham}.
In the AFQMC method, it is formally necessary to employ a split-operator method to separate $e^{-\tau \hat{H}_2}$ from $e^{-\tau (\hat H_1 + \hat H_2)}$, for example using first order split technique (Split-1)   
\begin{equation}
e^{-\tau (\hat H_1 + \hat H_2)} \approx  e^{-\tau \hat H_1} e^{-\tau \hat H_2}.
\end{equation}
For $e^{-\tau \hat H_2}$ the Hubbard-Stratonovich transformation is then used:
\begin{equation}
  e^{-\tau \hat H_2} = \frac{1}{N_w} \sum_w e^{-\tau \hat{\Heff}_{\mathrm{int}}^w} + \tau^2 \hat E_{\mathrm{HS}} + \mathcal{O}(\tau^3),   
\end{equation}
with the effective interaction introduced in Eq.~\eqref{eq:heffTOT}.
After some tedious algebra, we derive the leading error term of this transformation 
\begin{equation}
  \label{eq:hserror}
  \hat E_{\textrm{HS}} = \frac{1}{48} \sum_{g \neq g'} \bigl \{ \hat L_g \hat L_{g'}, [\hat L_{g'}, \hat L_{g}] \bigr \} + \frac{1}{24} \sum_{g \neq g'} \bigl [ \hat L_g, [\hat L_g, \hat L_{g'}^{2}] \bigr ],
\end{equation}
where $\{A, B\} = AB + BA$ and $[A,B] = AB - BA$.
The exact propagator is reproduced in the limit $N_w \to \infty$ and in case that $\hat L_g$ operators commute.
Note that in Eq.~\eqref{eq:hserror}, we only show the systematic errors remaining
after integrating over the random fields and not the statistical errors from the Monte
Carlo sampling.

After applying the Hubbard-Stratonovich transformation, one can choose to either evaluate the split propagator directly or refactorize it back to derive a full effective Hamiltonian
\begin{equation}
    \hat \Heff^w = \hat H_1 + \hat \Heff_{\mathrm{int}}^{w}
\end{equation}
and then apply any approximation to $e^{-\tau \hat \Heff^w}$.
In the subsequent analysis, we examine two split propagators where the matrix exponentials are evaluated by applying methods outlined in Sec.~\ref{sec:expm}.
Additionally, we examine two more propagators where the refactorized $e^{-\tau \hat \Heff^w}$ is approximated directly.

\subsection{Split-1 Propagator}
\label{sec:s1prop}
The simplest propagator splitting of the one-body Hamiltonian and the effective interaction is 
\begin{equation}
  \hat U_{\mathrm{S1}}(\tau) = \frac{1}{N_w} \sum_w e^{-\tau \hat H_1} e^{-\tau \hat{\Heff}_{\mathrm{int}}^w}.
\end{equation}
This has the advantage of reducing the number of necessary operations, however, 
the split-1 propagator leads to a sizable time-step error
\begin{equation}
    \hat U_{\mathrm{S1}}(\tau) = \hat U(\tau) + \tau^2 \hat E_{\mathrm{S1}} +  \mathcal{O}(\tau^3)
\end{equation}
with the leading term
\begin{equation}
    \hat E_{\textrm{S1}} = \frac{1}{2} \bigl[ \hat H_1, \hat H_2 \bigr] + \hat E_{\textrm{HS}},
\end{equation}
where $\hat E_{\textrm{HS}}$ is the error arising from the Hubbard-Stratonovich transformation (Eq.~\ref{eq:hserror}).
Since, the one-body propagator $e^{-\tau \hat H_1}$ remains constant during the simulation, we precomputed it exactly using matrix diagonalization.
Conversely, $e^{-\tau \Heff_{\mathrm{int}}^{w}}$ is generated anew for each walker and time step. Sec.~\ref{sec:expm} discusses efficient techniques for the evaluation of this term.

\subsection{Split-2 Propagator}
\label{sec:s2prop}
Split propagators are more accurate with second-order decomposition 
\begin{equation}
  \hat U_{\mathrm{S2}}(\tau) = \frac{1}{N_w} \sum_w e^{-\tau \hat H_1 / 2} e^{-\tau \hat{\Heff}_{\mathrm{int}}^w} e^{-\tau \hat H_1 / 2}.
\end{equation}
The advantage of this approach is that it does not introduce additional errors beyond the Hubbard-Stratonovich transformation to quadratic order
\begin{equation}
    \hat U_{\mathrm{S2}}(\tau) = \hat U(\tau) + \tau^2 
    \hat E_{\textrm{HS}}.
\end{equation}
Similar to the Split-1 propagator, the term $e^{-\tau \hat H_1 /2}$  is precomputed exactly in our implementation.

\subsection{Taylor Propagator}
\label{sec:taylorprop}
The simplest propagator within the AFQMC formalism is obtained by the Taylor expansion of the refactorized time propagator $e^{-\tau \Heff^w}$
\begin{equation}
  \hat U_{\mathrm{T}}(\tau) = \frac{1}{N_w} \sum_w e^{-\tau \hat \Heff^w} = \hat U(\tau) + \tau^2 \hat E_{\mathrm{T}} + \mathcal{O}(\tau^3).
\end{equation}
The leading error term in this expansion is
\begin{equation}
    \hat E_{\mathrm{T}} = \frac{1}{12} \sum_g \bigl[ [ \hat L_g, \hat H_1 ], \hat L_g \bigr] + \hat E_{\textrm{HS}},
\end{equation}
provided that the Taylor expansion is truncated at an order $k \geq 4$.

\subsection{Crank-Nicolson Propagator}
The Crank-Nicolson propagator replaces the matrix exponential by a linear problem
\label{sec:cnprop}
\begin{equation}
    \hat U_{\mathrm{CN}}(\tau) = \frac{1}{N_w}  \sum_w \bigl(1 + \tau \hat{\Heff}^w / 2\bigr)^{-1} \bigl(1 - \tau \hat{\Heff}^w / 2\bigr). 
\end{equation}
The corresponding leading error term is 
\begin{equation}
    \hat U_{\mathrm{CN}}(\tau) = \hat U(\tau) + \tau^2 \hat E_{\mathrm{CN}} + \mathcal{O}(\tau^3)
\end{equation}
with
\begin{equation}
    \hat E_{\textrm{CN}} = \frac{1}{4} \bigl \{ \hat H_1, \hat H_2 \bigr \} + \frac{1}{8} \sum_g \bigl[ [ \hat L_g, \hat H_1 ], \hat L_g \bigr] + \hat E_{\textrm{HS}}.
\end{equation}

\subsection{Summary}
In summary, the error terms specified above suggest that the Split-2 propagator is the optimal propagator within the AFQMC framework with the leading error term dictated by the Hubbard-Stratonovich transformation.
We will show below that this is also confirmed by numerical examples, albeit Taylor expansion will also yield 
excellent results almost on par with the Split-2 propagator. Conversely, the split-1 and the Crank-Nicolson propagator show significantly worse results.

\section{Exponential of the One-Body Operator within AFQMC}
\label{sec:expm}
To propagate according to the Split-2 propagator, we need to compute the action of the matrix exponential $e^{-\tau \hat{\Heff}_{\mathrm{int}}^w}$ and then apply it to the orbitals in $\ket{\Psi^w}$.
One could use exact matrix diagonalization\cite{GolubLoanMC,SaadKrylov} to compute $e^{-\tau \hat{\Heff}_{\mathrm{int}}^w}$;
however, this is computationally expensive especially because $\hat{\Heff}_{\mathrm{int}}^w$ is non-Hermitian (see Eq.~\ref{eq:heffTOT}).
While of limited practical utility, spectral decomposition still serves as a valuable benchmark tool.
We will use it to evaluate the accuracy of iterative methods.

An alternative is to use iterative methods to directly compute $e^{-\tau \hat{\Heff}_{\mathrm{int}}^w}\ket{\Psi^w}$.
All iterative methods involve $\hat{\Heff}_{\mathrm{int}}^w \ket{\Psi^w}$ operations that are the most expensive part of the procedure. 
Consequently, the optimal method for a fixed time step $\tau$ is the one that minimizes the number of $\hat{\Heff}_{\mathrm{int}}^w \ket{\Psi^w}$ operations.
To simplify the notation, we will use $\hat A = - \tau \hat{\Heff}_{\mathrm{int}}^w$ throughout the remainder of this section.

\subsection{Taylor Expansion}
\label{subsec:Taylor}
The Taylor expansion offers a simple and robust scheme to approximate matrix exponentials
\begin{equation}
    \label{eq:taylor}
    \textrm{taylor}_k\{A, \ket{\Psi} \} = \sum_{n=0}^{k} \frac{A^n}{n!} \ket{\Psi}~.
\end{equation}
$k$ is the order of the approximation and the number of $A\ket{\Psi}$ operations needed to approximate the exponential.
The scaling of the algorithm is dominated by $A\ket{\Psi}$ operations and scales as $kN^2N_e$.
Taylor expansion with $k=6$ is usually used in the AFQMC community.

\subsection{Chebyshev Expansion}
\label{sec:Cheb}
Ezer \emph{et al.}\cite{EzerCheb1984} demonstrated that, although the error terms of different polynomial expansions are of the same order, the Chebyshev polynomials $T_n(x)$ minimize the prefactor of the leading error term compared to other polynomial expansions.
They are conveniently defined using the following recurrence relation
\begin{align}
\begin{split}
  \label{eq:chebpoly}
  T_0(x) & = 1 \\
  T_1(x) & = x \\
  T_n(x) & = 2xT_{n-1}(x) - T_{n-2}(x).
\end{split}
\end{align}
The Chebyshev expansion is then formally given by
\begin{equation}
    \textrm{cheb}_k\{A, \ket{\Psi} \} = \sum_{n=0}^{k} c_n T_n(A) \ket{\Psi}~.
\end{equation}
For the matrix exponential, the expansion coefficients $c_n$ are
\begin{equation}
  c_n = \begin{cases}
         J_n(i) \quad & n=0 \\
         2i^nJ_n(-i) \quad & n > 0,
        \end{cases}
\end{equation}
where $J_n(z)$ are the Bessel functions of the first kind.
The $k$-th order Chebyshev expansion requires $k$ $A \ket{\Psi}$ operations and scales as $kN^2N_e$, same as the Taylor expansion.

\subsection{Krylov Subspace Projection}
\label{sec:Krylov}
The Krylov subspace $\mathcal{K}_k(A,\psi)$ for the matrix $A$ and an orbital vector $\psi$ is
\begin{equation}
    \label{eq:krylovsub}
    \mathcal{K}_k(A,\psi) = \text{span}\{\psi, A\psi,  A^2\psi, ...,  A^{k-1}\psi \}.
\end{equation}
In our case, $\psi$ is the $N$-dimensional vector of the orbital expansion coefficients in the Slater determinant $\ket{\Psi} = \ket{\psi_1, \psi_2,...,\psi_{N_e}}$.
For the non-Hermitian $A$, one can use the Arnoldi algorithm\cite{GolubLoanMC,SaadKrylov} to produce an orthonormal basis $B_k = [b_1, b_2,...,b_k]$ and the operator $A_k$ in the subspace 
\begin{equation}
    \label{eq:projham}
    A_k = B_k^{\dagger} A B_k.
\end{equation}
$A_k$ is a $k \times k$ upper Hessenberg matrix.\cite{GolubLoanMC}
$B_k$ is a $N \times k$ matrix with the first column $b_1 = \psi / ||\psi||$ and 
\begin{align}
    B_k^{\dagger}B_k & = \mathbb{1}_{k \times k},
    &
    B_k B_k^{\dagger} & \approx \mathbb{1}_{N \times N} .
\end{align}
Using this property, one can approximate the matrix exponential as follows:
\begin{align}
\begin{split}
    e^{A} \psi 
    & \approx B_k B_k^{\dagger}  e^{A} \overbrace{B_k e_1 ||\psi||}^{\psi} 
    \\
    & = B_k e^{A_k} e_1 ||\psi||,
\end{split}
\end{align}
where $e_1$ is the first unit vector in $\mathbb{C}^k$  and $B_k e_1 = b_1$.
The Krylov approximation to the matrix exponential of the order $k$ is then simply given by
\begin{equation}
    \label{eq:krylov}
    \textrm{krylov}_k \{ A, \psi \} = B_k e^{A_k} e_1 ||\psi||.
\end{equation}
Like the polynomial expansions, the $A\ket{\Psi}$ applications dominate the computational cost of the Krylov method with $kN^2N_e$ operations.
In addition, the Arnoldi algorithm scales as $kNN_e^2$ and one needs to compute the matrix exponential in subspace $e^{-\tau A_k}$.
The latter is typically not expensive because $k$ is very small compared to $N$; one computes it using matrix diagonalization or a higher-order polynomial expansion.

\subsection{Block-Krylov Subspace Projection}
\label{sec:BlockKrylov}
Contrary to the previous algorithm that expands each orbital in the Slater determinant individually, the block-Krylov projection simultaneously computes the action of the exponential operator on all orbitals.
As a consequence, the Krylov subspace increases by $N_e$ elements in each Arnoldi iteration, and $k$-th order Krylov subspace consists of $kN_e$ vectors.
The Krylov subspace is
\begin{equation}
    \label{eq:bkrylovsub}
    \mathcal{K}_k(A,\ket{\Psi}) = \text{span}\{\Psi, A\Psi,  A^2\Psi, ...,  A^{k-1}\Psi \},
\end{equation}
where $\Psi$ denotes the $N \times N_e$ dimensional matrix of the expansion coefficients.
The orthonormal basis $B_k = [\bar B_1, \bar B_2,...,\bar B_{k}]$ is now a $N \times kN_e$ matrix, where $\Psi = \bar B_1 R$, and can be computed using a QR decomposition.
Note that the initial orthogonalization of orbitals is necessary due to the imaginary-time propagation employed in AFQMC.
The projected operator $A_k$ is computed analogously to Eq.~\eqref{eq:projham} and is an upper block Hessenberg matrix of the size $kN_e \times kN_e$. 
The approximation of the exponential is similar to Eq.~\eqref{eq:krylov} 
\begin{equation}
    \label{eq:bkrylov}
    \textrm{bkrylov}_k \{ A, \ket{\Psi} \} = B_k e^{A_k} E_{N_e} R,
\end{equation}
where $E_{N_e}$ is the $k N_e \times N_e$ part of the $kN_e$-dimensional identity matrix.
For a given order $k$, the blocked Krylov method requires the same number of $A \ket\Psi$ operations as the non-blocked version.
The advantage of the blocked algorithm is that it may converge in fewer iterations because the space spanned by other orbitals is used to expand the exponential as well.
Furthermore, matrix-matrix multiplications often achieve more floating-point operations per second than matrix-vector multiplications.
The disadvantage is that the dimension of the Krylov subspace grows more rapidly so the cost of the Arnoldi operation and diagonalization increases more rapidly.

\subsection{Performance Comparison}
\label{sec:PerfComp}

\begin{figure}
    \centering
    \includegraphics[width=\columnwidth]{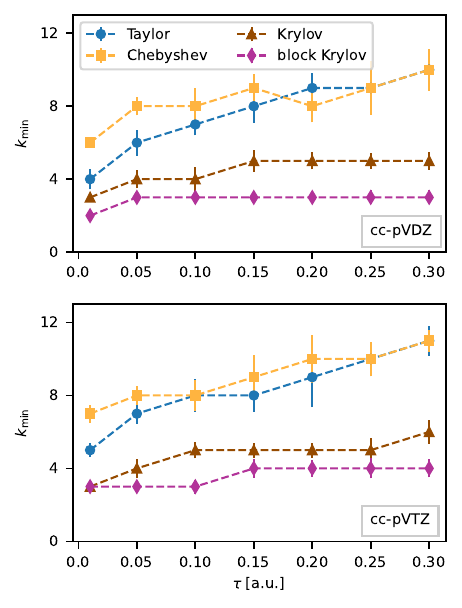}
    \caption{
    The number of $A \ket{\Psi}$ operations, i.e. the order of the method $k_{\mathrm{min}}$ required to achieve an accuracy of $\varepsilon =$ \qty{e-5}{\hartree} in the ph-AFQMC energy for different matrix-exponentiation methods.
    The local energy is averaged over 10 time steps and 2\,400 walkers.
    $k_{\mathrm{min}}$ is averaged over 22 systems.
    Note that the error bars correspond to the standard deviation of the average over the 22 systems and not the statistical fluctuations of the AFQMC simulation.
    Krylov methods (nonblocked, brown triangles; blocked, magenta diamonds) outperform the Taylor (blue circles) and the Chebyshev (yellow squares) expansion typically utilized in ph-AFQMC codes.
    The cc-pVDZ basis set (top) needs fewer $A \ket\Psi$ operations than the cc-pVTZ basis set (bottom).
    }
    \label{fig:expm}
\end{figure}

We compare the different matrix-exponential methods for a short ph-AFQMC simulation using 10 time steps and 2\,400 walkers and calculate the error in the ph-AFQMC total energy induced by the errors in calculating the matrix exponential. 
Then we determine the minimal number of $\hat A \ket{\Psi}$ operations $k_{\mathrm{min}}$ required to achieve a specified accuracy $\varepsilon =$ \qty{e-5}{\hartree}.
The exact diagonalization described at the beginning of the section yields the reference energies.
We average $k_{\mathrm{min}}$ over 22 different systems for time steps from $\tau =$ \qty{0.01}{\hartree^{-1}} to $\tau =$ \qty{0.3}{\hartree^{-1}} and for two different basis sets (cc-pVDZ and cc-pVTZ).
In Ref.~\onlinecite{prop_paper_data}, we provide the list of considered compounds, ranging from small molecules to benzene-like cyclic compounds.

Fig.~\ref{fig:expm} compares the different exponentiation methods and demonstrates that Krylov methods need less $\hat A \ket\Psi$ operations than the Taylor and the Chebyshev expansion.
Furthermore, $k_{\mathrm{min}}$ increases with the time step and this is more pronounced for the Taylor and the Chebyshev expansion than for the Krylov methods.
$k_{\mathrm{min}}$ is also more sensitive to the system for the Taylor and the Chebyshev expansion leading to larger error bars for these methods.
Hence, one needs to verify the order of the expansion for every system in particular for larger timesteps.
Finally, the larger cc-pVTZ basis set needs slightly larger $k_\mathrm{min}$ than the smaller cc-pVDZ basis set.

\section{Modified AFQMC Algorithm}
\label{sec:modalgo}
In this section, we present modifications of the commonly used AFQMC algorithm\cite{Motta2018AFQMCREVIEW,Shi2021RecentDevAFQMC,sukurmaHEAT2023} that allow large time steps in AFQMC and reliable time step extrapolation.

\subsection{Local-Energy Capping}
\label{subsec:A}
Inspired by Zen \emph{et al}\cite{Zen2016Boosting}, we implement the local energy capping as
\begin{equation}
\label{eq:enwin}
    \bar E_{\mathrm{L}}(\Psi) = \begin{cases}
        E_0 - \Delta E & \text{if} \quad E_{\mathrm{L}}(\Psi) < E_0 - \Delta E \\
        E_0 + \Delta E & \text{if} \quad E_{\mathrm{L}}(\Psi) > E_0 + \Delta E \\
        E_{\mathrm{L}}(\Psi) & \text{else.}
    \end{cases}
\end{equation}
Here, $E_{0}$ is the best estimate of the ground-state energy and the energy window is $\Delta E = \frac{1}{2} \sqrt{\frac{N_e}{\tau}} + \sqrt{N_e \tau}$.
This window increases with the number of electrons as opposed to $\Delta E = \sqrt{\frac{2}{\tau}}$ typically used in AFQMC.\cite{PurwantoSiTin2009}
We also add a term proportional to $\sqrt{\tau}$ to avoid excessively strong capping for large time steps.
The energy capping introduces a bias, potentially affecting the accuracy of the time step extrapolation.
To strike a balance, we opt to cap the local energy to control its variance only---a parameter far more sensitive to rare events than the mean value.
The energy capping is usually applied for the hybrid energy (Eq.~\eqref{eq:hyben}), too.
We turn off hybrid energy capping because it strongly affects the time-step errors. 

\subsection{Reweighting Factor Capping}
To compensate for the lack of hybrid energy capping, we restrict the reweighting factor in Eq.~\eqref{eq:wupdate}
\begin{equation}
   \overline{e^{-\tau(E_{\mathrm{H}}-E_0)}} = \begin{cases}
        e^{-\tau(E_{\mathrm{H}}-E_0)} & \text{if} \quad e^{-\tau (E_{\mathrm{H}}- E_0)} < 10 \\
        0   & \text{else~.}
    \end{cases}
\end{equation}
This is as effective as the hybrid-energy capping and does not impact the time-step errors.

\subsection{Force-Bias Capping}
Force-bias capping is essential for stable AFQMC simulations.
We recommend capping the force bias with
\begin{equation}
    \bar f_g = \begin{cases}
        f_g & \text{if} \quad |f_g| < 1 \\
        0   & \text{else~.}
    \end{cases}
\end{equation}
to minimize the fluctuations and time-step errors.
We find that removing the force-bias components with large absolute values
is more reliable for large time steps than the conventional choice of capping them at an absolute value of 1.

\subsection{No Energy Mixing}
\label{subsec:D}
Many implementations modify Eq.~\eqref{eq:wupdate} to include a mixture $E_\mathrm{H}^\mathrm{mix}$ 
 of the current and previous hybrid energy
\begin{equation} 
    W_{k+1}^w e^{i \theta_{k+1}^w} = W_{k}^w e^{i \theta_{k}^w} \; e^{-\tau (E_{\textrm{H}}^{\mathrm{mix}} - E_0)}, 
\end{equation}
However, we find that energy mixing deteriorates the time-step extrapolation, so we do not use it in this work.

\subsection{Accurate Matrix Exponentials}
Reliable time-step extrapolation requires minimizing remaining errors caused by approximating the exponential of the effective interaction $\hat{\Heff}_{\mathrm{int}}$.
We recommend using either the Krylov projection method with $k=5$ or the Block-Krylov method with $k=4$ for this purpose (see Sec.~\ref{sec:expm}).

\subsection{Size Consistency Errors}
\begin{figure}
    \centering
    \includegraphics[width=\columnwidth]{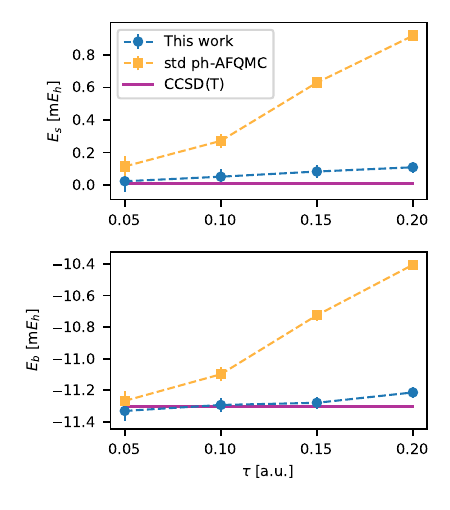}
    \caption{
    (Top) The size-consistency error $E_\mathrm{s}$ and (bottom) the binding energy $E_\mathrm{b}$ for a CH$_4$-H$_2$O dimer using a cc-pVDZ basis set.
    We compare the standard ph-AFQMC algorithm (yellow squares), our modified algorithm (blue circles), and CCSD(T) results (magenta line). 
    }
    \label{fig:sizecons}
\end{figure}

To quantify the effect of the listed modifications, we test the size consistency of the algorithms similar to Zen \emph{et al}. \cite{Zen2016Boosting}
We consider a CH$_4$-H$_2$O dimer at a separated C-O distance of \qty{11.44}{\angstrom} using the cc-pVDZ basis set.
We measure the size-consistency error using
\begin{equation}
   E_\mathrm{s} = E_{\mathrm{CH}_4-\mathrm{H}_2\mathrm{O}}^{\mathrm{separated}} - E_{\mathrm{CH}_4} - E_{\mathrm{H}_2\mathrm{O}}.
\end{equation}
We expect that $E_{\mathrm{s}}$ is less than \qty{0.01}{\milli \hartree} for size-consistent methods.
$E_s$ is small for both algorithms at small time steps ($\tau = 0.05$~\si{\hartree^{-1}}, see Fig.~\ref{fig:sizecons}(top)).
However, the standard algorithm exhibits sizable errors at larger time steps ($\approx$ \qty{0.4}{\milli \hartree} at $\tau =$ \qty{0.20}{\hartree^{-1}}), whereas the modified algorithm is size-consistent for all time steps.
These size-consistency errors directly impact the binding energy 
\begin{equation}
   E_\mathrm{b} = E_{\mathrm{CH}_4-\mathrm{H}_2\mathrm{O}}^{\mathrm{bonded}} - E_{\mathrm{CH}_4} - E_{\mathrm{H}_2\mathrm{O}},
\end{equation}
where the bonded C-O distance is \qty{0.63}{\angstrom}.
Fig.~\ref{fig:sizecons}(bottom) shows similar-size deviations of the energy when the dimer is bound.
Hence, the standard algorithm requires much smaller time-steps to achieve the same level of accuracy as the algorithm presented in this work.

\subsection{Time-Step Extrapolation of ph-AFQMC Energies}
\label{sec:tauext}

\begin{figure}
    \centering
    \includegraphics[width=\columnwidth]{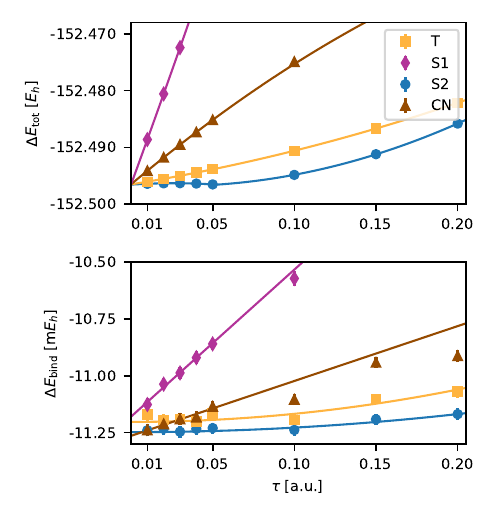}
    \caption{
    Time-step errors in the total energy (top) and the binding energy (bottom) of the 2C$_\mathrm{S}$ cluster for different AFQMC propagators.
    Time steps $<$ \qty{0.05}{\hartree^{-1}} are used for the linear extrapolation of the binding energy with the Crank-Nicolson (brown triangles) and the Split-1 (magenta diamonds) propagators, while all time steps are used for the quadratic extrapolation with the Taylor (yellow squares) and the Split-2 (blue circles) propagator.
    All calculations use the frozen-core approximation and the cc-pVDZ basis set.
    }
    \label{fig:tau_ext}
\end{figure}

Generally, the time-step errors in the total energy follow a quadratic relation
\begin{equation}
    \label{eq:square_fit}
    E(\tau) = E_0 + \alpha \tau + \beta \tau^2
\end{equation}
for all propagators.
The quadratic term directly emerges from the analysis of the propagators (Sec.~\ref{sec:integrators}), whereas the linear term is attributed to perturbations introduced in the Monte Carlo algorithm to maintain the stability of the random walk (Sec.~\ref{sec:modalgo}.A$\, - \,$\ref{sec:modalgo}.D).
The top plot in Fig.~\ref{fig:tau_ext} illustrates the time-step errors for the absolute ph-AFQMC energies of the 2C$_{\mathrm{S}}$ water dimer with the cc-pVDZ basis set.
Notably, the Split-2 propagator exhibits the smallest time-step errors predominantly characterized by the quadratic term.
The Taylor propagator shows similar behavior with slightly larger time-step errors.
Conversely, both the Crank-Nicolson and the Split-1 propagator show larger time-step errors primarily governed by the linear term.

The bottom plot of Fig.~\ref{fig:tau_ext} displays the time-step extrapolation for relative energies, specifically the binding energy of the 2C$_{\mathrm{S}}$ water dimer.
For relative energies, time-step errors cancel by 1--2 orders of magnitude compared to the absolute energies.
Again, the Split-2 and Taylor propagators show the smallest time-step errors with vanishingly small linear terms.
For these propagators and this system, the time-step errors are so small ($< $ \qty{0.1}{\milli \hartree}) that extrapolation is not necessary and the binding energy could be determined by a single calculation.
Conversely, for the Split-1 and Crank-Nicolson propagators, the linear term remains significant, leading to larger time-step errors.
Based on these observations, we recommend a general rule for extrapolation of the binding energies: use 
\begin{equation}
\label{eq:pure_square_fit}
    E(\tau) = E_0 + \beta \tau^2
\end{equation}
 for propagators where quadratic errors dominate (Split-2 and Taylor), and 
\begin{equation}
    E(\tau) = E_0 + \alpha \tau    
\end{equation}
for those propagators with the predominant linear term (Split-1 and Crank-Nicolson). 

\section{Applications and Discussion}
\label{sec:results}

In the following three subsections, we
(i) compare accurate ph-AFQMC energies obtained with small time steps\cite{sukurmaHEAT2023} to extrapolating the energies from large time-steps,
(ii) extrapolate the binding energies of small water clusters in the heavy-augmented double-zeta basis set and in the CBS limit, and
(iii) calculate the equilibrium bond length of the N$_2$ dimer using large time-step ph-AFQMC calculations.

\subsection{HEAT Set Total Energies}
\label{sec:heat}
We calculate the total energies for the molecules in the HEAT set\cite{Tajti2004Heat} using the Split-2 propagator and time-step extrapolation.
We include Cholesky vectors up to a threshold of \qty{e-6}{\hartree} and sample with the Hartree-Fock trial determinant.
We use Dunning's cc-pVDZ basis set and the frozen-core approximation (1$s$ shell for B-F atoms).
To verify the reliability of the time-step extrapolation, we utilized 4 equidistant time steps ranging between \qty{0.05}{\hartree^{-1}} and \qty{0.20}{\hartree^{-1}}.
We perform each ph-AFQMC simulation using 2\,400 walkers, with 2\,000 equilibration steps, and 10\,000 time steps during the sampling phase.
To have accurate reference values for comparison, we perform small time-step ph-AFQMC simulations with $\tau =$ \qty{0.002}{\hartree^{-1}}, 4\,800 walkers, 100\,000 equilibration steps, and 400\,000 sampling steps.
This setup guarantees roughly identical statistical errors for the small time-step and extrapolated energies, but the CPU time to obtain extrapolated energies is reduced by a factor of 20 using large time steps. 
All calculations are performed using the QMCFort code and a single CPU node (dual-socket Intel Skylake Platinum 8174) per system.

\begin{figure}
  \centering
  \includegraphics[width=\columnwidth]{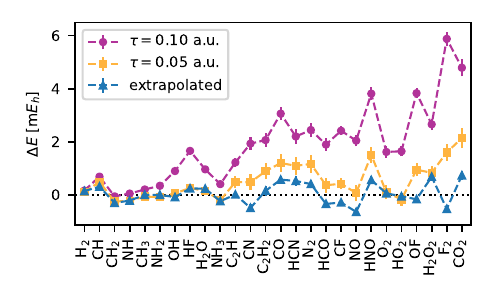}
  \caption{
    Comparison of time-step extrapolated and large time-step energies ($\tau =$ \qty{0.05}{\hartree^{-1}} and $\tau =$ \qty{0.10}{\hartree^{-1}}) against reference values for 26 molecules in the HEAT set.
    Reference values are obtained using small time-step ph-AFQMC with $\tau =$  \qty{0.002}{\hartree^{-1}}.
    All computations use the cc-pVDZ basis set and the frozen-core approximation.
  }
  \label{fig:heat}
\end{figure}

Fig.~\ref{fig:heat} compares the large time-step and extrapolated ph-AFQMC energies with the reference ones at $\tau =$ \qty{0.002}{\hartree^{-1}}.
The ph-AFQMC energies at $\tau =$ \qty{0.10}{\hartree^{-1}} exhibit sizable time-step errors with a mean error of \qty{1.89}{\milli \hartree}.
The time-step errors lead to a consistent under-correlation.
Reducing the time step to $\tau =$ \qty{0.05}{\hartree^{-1}} reduces the mean error to \qty{0.53}{\milli \hartree} (already a small fraction of the desired chemical accuracy of \qty{1.59}{\milli \hartree}).
The extrapolated energies are accurate with a mean absolute error of \qty{0.31}{\milli \hartree}.
The mean signed error of \qty{0.06}{\milli \hartree} shows that the extrapolated energies are not systematically biased compared to the reference small time-step energies.

\subsection{Water Clusters}
\label{sec:water}
As a second example, we apply time-step extrapolation to determine the binding energies of the water clusters.
We utilize the Split-2 propagator at three distinct time steps ($\tau = 0.10$, $\tau = 0.15$, and $\tau =$ \qty{0.20}{\hartree^{-1}}) to extrapolate the binding energies using Eq.~\eqref{eq:pure_square_fit}.
To make reference calculations with a small time step possible, we employ the heavy-augmented double-zeta basis set.
Furthermore, we obtain the CBS limit using the aug-cc-pV$N$Z ($N = \text{D}, \text{T}, \text{Q}$) basis sets and a 4-5 inverse polynomial extrapolation scheme\cite{TemesloWaterClusters2011}
\begin{equation}
    E_{\mathrm{CBS}} = E_{N} + \frac{b}{(N + 1)^4} + \frac{c}{(N+1)^5}
\end{equation}
due to its superior behavior compared to the standard inverse cubic extrapolation scheme.
Here, $E_{\mathrm{CBS}}$ is the energy in the complete basis set, $N$ is the largest angular momentum in the aug-cc-pV$N$Z basis set ($N=2-4$ for $N=\mathrm{D,T,Q}$), and $E_N$ is the corresponding energy.

Each ph-AFQMC calculation is performed using 2\,400 walkers and 50\,000 time steps, except for the water molecule, where we used 9\,600 walkers and 200\,000 time steps to minimize binding energy errors.
The computational setup is described in more detail in our previous work.\cite{sukurmaHEAT2023}
As an illustrative example, the largest calculation was performed on the 5CYC cluster using the aug-cc-pVQZ basis set, comprising 855 orbitals, 40 electrons, and 5685 Cholesky vectors.
Executed on 4 CPU nodes (dual-socket AMD Epyc 7713), this calculation took 8 days to complete (average performance of \qty{40}{GFLOPS / core}).

\begin{table}
\caption{\label{tab:clusters_hvdz} Binding energies of the four most stable water clusters calculated using a heavy-augmented cc-pVDZ basis set and all-electron wavefunctions. MP2, CCSD(T), small time-step ph-AFQMC, and time-step extrapolated ph-AFQMC values (ph-AFQMC(x)) are given in \unit{\kilo \cal / \mole}.}
\begin{ruledtabular}
\begin{tabular}{l*{4}{d}}
Cluster             & \multicolumn{1}{c}{MP2} & \multicolumn{1}{c}{CCSD(T)} & \multicolumn{1}{c}{ph-AFQMC \cite{sukurmaHEAT2023}} & \multicolumn{1}{c}{ph-AFQMC(x)} \\ \hline
2Cs                 & -5.22    & -5.18   & -5.17(5)    & -5.24(3)    \\
3UUD                & -15.83   & -15.62  & -15.67(9)   & -15.81(4)   \\
4S4                 & -28.36   & -27.87  & -28.12(10)  & -28.14(5)  \\
5CYC                & -37.48   & -36.78  & -37.14(28)  & -37.27(6)  \\ 
\end{tabular}
\end{ruledtabular}
\end{table}

Table \ref{tab:clusters_hvdz} presents time-step extrapolated ph-AFQMC binding energies for water clusters and compares them with MP2, CCSD(T), and small time-step ($\tau =$ \qty{0.01}{\hartree^{-1}}) ph-AFQMC values reported in our previous work.\cite{sukurmaHEAT2023}
The time-step extrapolated binding energies are within the error bars in agreement with the small time-step ph-AFQMC values and also agree with MP2 and CCSD(T) values within chemical accuracy. 
Compared to the small time-step ph-AFQMC calculations, the total number of samples is reduced by a factor of 2, while concurrently reducing the statistical errors, too.
We used a simple expression to estimate the statistical error $\sigma_{\mathrm{x}}$ for the extrapolated energies
\begin{equation}
\label{eq:staterr}
    \sigma_{\mathrm{x}} = \frac{\sqrt{\sum_{i=1}^{N_{\tau}} \sigma_i^2}}{N_{\tau}}~,
\end{equation}
where $N_{\tau}$ is the number of time steps used for the extrapolation and $\sigma_i$ denotes the estimated 1$\sigma$ error for each of the time steps.  

\begin{table}
\caption{\label{tab:clusters_cbs} CBS binding energies of the four most stable water clusters estimated using aug-cc-pV(D,T,Q) basis sets and 4-5 inverse polynomial extrapolation scheme.
Frozen-core approximation is employed.
MP2, CCSD(T), and time-step extrapolated ph-AFQMC values in \unit{\kilo \cal / \mole} are reported.}
\begin{ruledtabular}
\begin{tabular}{l*{4}{d}}
Cluster             & \multicolumn{1}{c}{MP2 \cite{TemesloWaterClusters2011}} & \multicolumn{1}{c}{CCSD(T) \cite{TemesloWaterClusters2011}} & \multicolumn{1}{c}{ph-AFQMC(x)} \\ \hline
2Cs                 & -5.00    & -5.03   & -4.98(3)    \\
3UUD                & -15.72   & -15.70  & -15.82(5)   \\
4S4                 & -27.64   & -27.43  & -27.53(8)  \\
5CYC                & -36.38   & -36.01  & -36.01(10)  \\ 
\end{tabular}
\end{ruledtabular}
\end{table}

Table \ref{tab:clusters_cbs} lists the MP2, CCSD(T),\cite{TemesloWaterClusters2011} and our time-step extrapolated ph-AFQMC binding energies in the CBS limit.
Small time-step ph-AFQMC calculations are not possible in the CBS limit due to the large computational cost.
The mean absolute error between our results and MP2 amounts only to \qty{0.15}{\kilo \cal / \mol}.
The ph-AFQMC results deviate even less from CCSD(T) with a mean absolute error of \qty{0.07}{\kilo \cal / \mol}.

We estimate the statistical errors of the extrapolated binding energies using Eq.~\eqref{eq:staterr}.
Moreover, for the statistical errors after the basis set extrapolation, we used the corresponding errors observed at the aug-cc-pVQZ basis set. 
We note that the CCSD(T) values are approximated by adding the difference between CCSD(T) and MP2 at the aug-cc-pVDZ basis set to the CBS MP2 binding energies.

\begin{figure}
    \centering
    \includegraphics[width=\columnwidth]{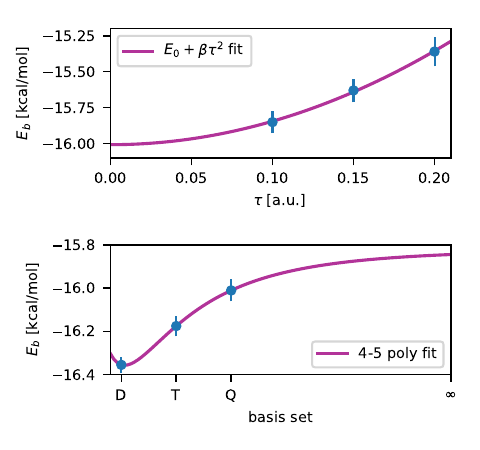}
    \caption{
    (Top) Time-step extrapolation of the binding energy for the 3UUD water cluster at the aug-cc-pVQZ basis set. 
    Eq.~\eqref{eq:pure_square_fit} is used to estimate the zero time-step limit.
    (Bottom) Basis set extrapolation of the binding energy for the 3UUD water cluster using aug-cc-pV(D,T,Q) basis sets and 4-5 inverse polynomial extrapolation technique.
    }
    \label{fig:cluster}
\end{figure}

As an example, we show the time-step extrapolation of the binding energy for the 3UUD water cluster at the aug-cc-pVQZ basis set in Fig.~\ref{fig:cluster}(top).
Fig.~\ref{fig:cluster}(bottom) illustrates the basis set extrapolation for the same cluster.

\subsection{Equilibrium Geometry of the N$_2$ Dimer}
\label{sec:n2}
Unlike the total energies, properties such as energy differences, bond lengths, and lattice constants exhibit significantly lower sensitivity to time-step errors.
Using the N$_2$ molecule, we demonstrate that accurate bond lengths are successfully obtained with single-point, large time-step ph-AFQMC calculations.
Using the aug-cc-pVQZ basis set and a RHF trial wave function, we calculate ph-AFQMC energies for five equidistant bond lengths ranging from \qty{0.9}{\angstrom} to \qty{1.3}{\angstrom}.
We compare the results with a small time step of \qty{0.004}{\hartree^{-1}} to a larger one of \qty{0.1}{\hartree^{-1}}.
Fitting total energies to the Morse potential energy curve
\begin{equation}
    \label{eq:morse}
    E(R) = E_0 + D (1 - e^{-a(R-R_0)})^2~,
\end{equation}
determines the equilibrium bond length $R_0$, the dissociation energy $D$, and the parameter $a$ controlling the width of the potential.
\begin{figure}
    \centering
    \includegraphics[width=\columnwidth]{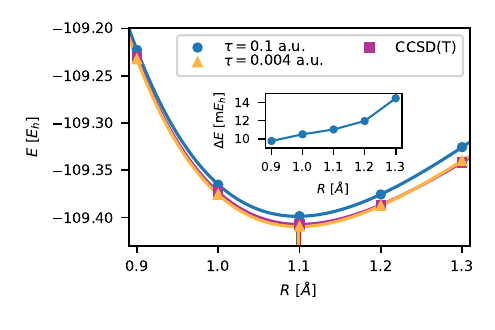}
    \caption{
    The potential energy curves $E$ obtained for CCSD(T), ph-AFQMC with $\tau =$ \qty{0.004}{\hartree^{-1}}, and ph-AFQMC at $\tau =$ \qty{0.1}{\hartree^{-1}} using aug-cc-pVQZ basis set. 
    Five equidistant bond lengths in the range from \qty{0.9}{\angstrom} to \qty{1.3}{\angstrom} are used to fit the Morse potential.
    The equilibrium bond lengths agree closely with the experimental value.
    Furthermore, the inset illustrates the behavior of ph-AFQMC time-step errors $\Delta E$ with increasing bond length.
    }
    \label{fig:n2eqgeo}
\end{figure}

\begin{table}
\caption{\label{tab:bondlength}
Equilibrium bond lengths of the N$_2$ dimer for different levels of theory: CCSD(T), ph-AFQMC with small time step, and the ph-AFQMC at large time step.
}
\begin{ruledtabular}
\begin{tabular}{ll}
\multicolumn{1}{c}{Method} & \multicolumn{1}{c}{$R_0$ [~\AA{}]} \\ \hline
CCSD(T)                                          & 1.100(1) \\
ph-AFQMC @ \SI{0.004}{\hartree^{-1}}             & 1.099(2) \\
ph-AFQMC @ \SI{0.1}{\hartree^{-1}}               & 1.098(2) \\  \hline
Experiment                                       & 1.098  \cite{HuberHerzberg}   \\
\end{tabular}
\end{ruledtabular}
\end{table}

Figure \ref{fig:n2eqgeo} shows both ph-AFQMC and the reference CCSD(T) potential energy curves.
The absolute CCSD(T) and small time step ph-AFQMC energies agree very well, however, the large time step ph-AFQMC energies are shifted by \qty{10}{\milli \hartree}--\qty{14}{\milli \hartree} (inset in Fig~\ref{fig:n2eqgeo}).
The slight increase of the ph-AFQMC time-step errors with increasing bond length may result from the deteriorating quality of the trial RHF wave function.  
Nevertheless, the observed change in time-step errors is not large enough to compromise the accuracy of the predicted bond length;
the predicted bond lengths listed in Table~\ref{tab:bondlength} are nearly identical and align well with the experimental value within statistical accuracy.
Using large-time step ph-AFQMC reduces the computational time for each calculation by a factor of $\tau_{\mathrm{L}}/\tau_{\mathrm{S}}=25$.

\section{Conclusion}
\label{sec:conclusion}
We implemented modifications of the ph-AFQMC algorithm in the QMCFort code to reduce the size-consistency errors at large time steps.
The modifications are mainly related to the control of rare events and the weight update procedure.
Addressing size-consistency errors is crucial as they have a consequential impact on the accuracy of binding energies. 
Using the CH$_4$-H$_2$O dimer (Fig.~\ref{fig:sizecons}) we showed that the modified approach markedly diminishes size-consistency errors across all time steps. 

Employing large time steps required better control of the errors in the matrix exponentiation of the one-body operators.
Utilizing a diverse set of 22 different systems using Dunning's cc-pVDZ and cc-pVTZ basis sets, we demonstrated that the common sixth-order Taylor expansion may prove insufficient.
Larger time steps require on average up to 10--12 $\hat{\Heff} \ket{\Psi}$ operations for adequate accuracy. 
In contrast, Krylov methods--- particularly a blocked variant ---are more robust, with negligible time-step errors while employing merely 4 $\hat{\Heff} \ket{\Psi}$ operations even for large time steps.
The block-Krylov method is also less sensitive to system and basis set choice than the Taylor expansion.
Moreover, it introduces negligible computational overhead, especially for larger systems, making it a promising alternative.

While all AFQMC propagators yield consistent results for small time steps, we have shown that the Split-2 propagator is the optimal propagator within the AFQMC formalism, with the leading error term dictated by the Hubbard-Stratonovich transformation.
We demonstrated that reliable time-step extrapolation is possible for all propagators via a second-order polynomial fit. 
Notably, the Split-1 and Crank-Nicolson propagators exhibit significant linear time-step errors compared to the Taylor and Split-2 propagators that are dominated by quadratic errors.
For the HEAT set molecules, using a Split-2 propagator reduced the computational cost by an order of magnitude while retaining almost the same accuracy (mean absolute error of \qty{0.31}{\milli \hartree}) as the small time-step reference.
For water clusters, the time-step extrapolated binding energies at the heavy-augmented double-zeta basis set agree well with those obtained by small time-step calculations.
Similarly, the complete basis-set binding energies are in excellent agreement with the reference CCSD(T) values (root-mean-square error of \qty{0.08}{\kilo \cal / \mol}).

Finally, we showed that certain observables, such as the equilibration bond length, are much less sensitive to time-step errors than total energies.
Using the N$_2$ dimer as an example, we demonstrated that the ph-AFQMC calculation with $\tau =$ \qty{0.1}{\hartree^{-1}} yields the same bond length as a calculation with $\tau =$ \qty{0.002}{\hartree^{-1}}, while reducing the computational cost by a factor of 25.

In conclusion, we highlighted the ability to employ larger time steps in ph-AFQMC, resulting in an order of magnitude speedup compared to standard ph-AFQMC calculations.
We plan to validate the efficacy of large time-step ph-AFQMC on a wider range of systems where accurate small time-step computations are very expensive. 

\section{Acknowledgments}
Funding by the Austrian Science Foundation (FWF) within the project P 33440 is gratefully acknowledged. All calculations were performed on the VSC4 / VSC5 (Vienna scientific cluster). 

\section*{Author declarations}
\subsection*{Conflict of Interest}
The authors have no conflicts to disclose.

\section*{Data availability}
The additional data that support the findings of this study are available online (see Ref.~\onlinecite{prop_paper_data}).

\bibliographystyle{aapmrev4-1}
\bibliography{bibliography}         

\end{document}